\documentclass[seceq]{ptptex}
\usepackage{wrapft}
\usepackage{graphicx}

\newcommand{\ket}[1]{| {#1} \rangle}     
\newcommand{\kett}[1]{| {#1} \rangle\!\rangle}     
\newcommand{\kket}[1]{|\!| {#1} \rangle}     
\newcommand{\wtilde}[1]{\widetilde{#1}} 

\newcommand{\kkett}[1]{|| {#1} \rangle\!\rangle}     
\def\beq{\begin{eqnarray}}
\def\eeq{\end{eqnarray}}
\def\bsub{\begin{subequations}}
\def\esub{\end{subequations}}
\def\b{\begin{equation}}

\title{
On the Exact Eigenstates and the Ground States Based on the Boson Realization 
for Many-Quark Model with $su(4)$ Algebraic Structure
}

\author{
Yasuhiko {\sc Tsue},$^{1}$ Constan\c{c}a {\sc Provid\^encia},$^{2}$ 
Jo\~ao da {\sc Provid\^encia}$^{2}$  and Masatoshi {\sc Yamamura}$^{3}$  
}

\inst{
$^{1}$Physics Division, Faculty of Science, Kochi University, Kochi 780-8520, 
Japan\\
$^{2}$Departamento de F\'{i}sica, Universidade de Coimbra, 3004-516 Coimbra, 
Portugal\\
$^{3}$Department of Pure and Applied Physics, 
Faculty of Engineering Science,\\
Kansai University, Suita 564-8680, Japan
}

\recdate{
\today
}

\abst{
Based on the boson realization, 
the ground state and phase structure are investigated in a many-quark 
model with algebraic structure developed in our previous paper, 
in which the two-body pairing and particle-hole type interactions are 
active. 
The physical interpretation of the exact eigenstates obtained in the boson 
realization in our previous paper 
is given in a many-quark model with $su(4)$ algebraic structure. 
}

\begin{document}

\maketitle

\section{Introduction}

Recently, the exact eigenstates and energy eigenvalues were obtained 
in a many-quark model, in which the two-body pairing and the particle-hole 
type interactions are active, by present authors in Ref.\citen{1}, 
which is hereafter referred to as (I) in this paper. 
This model was called the modified Bonn model because, 
if the only pairing interaction is included, this model is 
reduced to the Bonn model developed by Petry et. al.\cite{Petry} 
It is pointed out in (I) that the 
original Bonn model has the $su(4)$ symmetry. 
In the original Bonn model, the colored state is 
energetically favorable while the color neutral triplet formation of quarks 
is realized as a nice feature of the model.\cite{Pittel} 
Thus, this model leads to the pairing instability and the ground state 
is color superconducting state at any time. 

However, the dynamics of quarks are govened by 
the quantum chromodynamics (QCD) which posseses a 
color $su(3)$ symmetry. 
Therefore, there is a room that 
the original Bonn model can be extended to the 
$su(3)$ symmetric model, not $su(4)$, by introducing the $su(4)$ symmetry 
breaking term. 

In (I), we have formulated the modified Bonn model with color $su(3)$ 
symmetry introducing the particle-hole type interaction which breaks the 
$su(4)$ symmetry. 
In that paper, the exact eigenstates and the exact energy eigenvalues 
have been 
obtained in the boson space by using the Schwinger boson representation. 
In the triplet formation of quarks, the energy of the colored state 
is smaller than that of the color neutral triplet states if the particle-hole 
type interaction 
is switched off, namely, in the original Bonn model. 
However, the energy of the colored triplet states sifts up 
as the coupling strength of the particle-hole type interaction is increasing, 
while the energy of the color neutral triplet state is not changed. 
Thus, beyond a certain critical value of the coupling strength, 
the color neutral triplet state is favorable energetically compared with 
the colored triplet state. 
As is similar to the case of the triplet formation, 
in the pairing states, the energy of colored pairing states 
is energetically favorable. But, the energy of these states increses 
as the coupling strength of the particle-hole type interaction 
increases, as is seen in the case of the colored triplet states in 
the triplet formation. 
Further, in Ref.\citen{2}, which is referred to as (II) in this paper, 
the eaxct eigenstates are described in a unfied manner. 
However, in (I) and (II), it is not investigated which state, namely, 
the triplet state or the pairing state, is energetically favorable 
or not. 

In this paper, as one of important purposes rested in our previous papers, 
we investigate the ground state in the modified Bonn model. 
Following the previous papers (I) and (II), we describe the exact 
eigenstates and energy eigenvalues of this many-quark model 
in the boson space by using the boson realization. 
We will give the phase diagram in this many-quark model under 
a certain parameterization. 

This paper is organized as follows: 
In the next section, we recapitulate the exact eigenstates and the energy 
eigenvalues for this many-quark model in the boson realization. 
In \S 3, the physical interpretation of the exact eigenstates obtained 
in the boson representation of many-quark model is discussed in some cases 
by using the correspondences with the original fermion states. 
In \S 4, the ground state and the phase structure 
in this modified Bonn model are investigated. 
The last section is devoted to the discussions and concluding remarks.

\section{Recapitulation of the exact eigenstates and the energy eigenvalues 
for many-quark model with $su(4)$ algebraic structure}

In this section, we recapitulate the basic ingredients of the 
many quark model with $su(4)$ algebraic structure, especially, 
of the energy eigenvalues and eigenstates for the model Hamiltonian 
following to (I). 

\subsection{The model}

The model is formulated in terms of the generators of 
the $su(4)$ algebra constructed by the bilinear forms of 
the quark creation and annnihilation 
operators. 
The color quantum numbers are specified as $i=1,2,3$. 
Each color state has the degeneracy 
$2\Omega$. Here, $2\Omega=2j_s+1$ and $j_s$ is a half integer. 
Thus, the maximum total quark number is $6\Omega$. 
An arbitrary single-particle state is specified as $(i,m)$ with 
$m=-j_s,\ -j_s+1,\cdots,j_s-1,\ j_s$, and is created and 
annihilated by the fermion operators $c^*_{im}$ and $c_{im}.$ 
For simplicity, we neglect the degrees of freedom related to the 
isospin. 
We define 
the following bilinear forms for the fermion creation and annihilation 
operators:
\begin{eqnarray}\label{2-1}
&&\tilde S^1=\sum_m c^*_{2m}c^*_{3\tilde m},\quad \tilde
S^2=\sum_mc^*_{3 m}c^*_{1\tilde m},\quad \tilde
S^3=\sum_mc~^*_{1m}c^*_{2\tilde m},\nonumber\\
&&\nonumber\tilde S_1^2=\sum_mc^*_{2m}c_{1m},\quad \tilde
S_2^3=\sum_mc^*_{3m}c_{2m},\quad \tilde
S_3^1=\sum_mc^*_{1m}c_{3m}, \nonumber\\
&&\nonumber\tilde
S^1_1=\sum_m(c^*_{2m}c_{2m}+c^*_{3m}c_{3m})-2\Omega,\quad
\tilde S^2_2=\sum_m(c^*_{3m}c_{3m}+c^*_{1m}c_{1m})-2\Omega,\\
&&\nonumber\tilde
S^3_3=\sum_m(c^*_{1m}c_{1m}+c^*_{2m}c_{2m})-2\Omega,\quad\tilde
S_1=(\tilde S^{1})^*,\quad \tilde S_2=(\tilde S^{2})^*,\quad \tilde
S_3=(\tilde S^{3})^*,\\&& \tilde S^1_2=(\tilde S^2_1)^*,\quad \tilde
S^2_3=(\tilde S^3_2)^*,\quad\tilde S^3_1=(\tilde S^1_3)^*\label{1} \ .
\end{eqnarray}
Here, $c^*_{i\tilde m}=(-1)^{j_s-m}c^*_{i,~-m}.$ 
The operators in 
the definition (\ref{2-1}) are generators of the $su(4)$ algebra: 
\begin{eqnarray}\label{2-2}
&&
\tilde S^*_i=\tilde S^i,\quad (\tilde S^i_j)^*=\tilde S^j_i,\quad
\nonumber [\tilde S^i,\tilde S^j]=0,\quad  [\tilde S^i,\tilde
S_j]=\tilde S^j_i ,\\&&  [\tilde S^j_i,\tilde S^k]=\delta_{ij}\tilde
S^k+\delta_{jk}\tilde S^i, \quad  [\tilde S^j_i,\tilde
S^k_l]=\delta_{jl}\tilde S^k_i-\delta_{ik}\tilde S^j_l.
\end{eqnarray}
The fermion number operators with $i=1,2,3$ read, respectively, 
\begin{subequations}\label{2-3}
\begin{eqnarray}\label{2-3a}
& &\wtilde N_1=\Omega-\frac{1}{2}
\left(\wtilde S^1_1-\wtilde S^2_2-\wtilde
S^3_3\right),\quad 
\wtilde N_2=\Omega-\frac{1}{2}\left(\wtilde S^2_2-\wtilde
S^3_3-\wtilde S^1_1\right),\nonumber\\ 
& &\wtilde N_3=\Omega-\frac{1}{2}\left(\wtilde
S^3_3-\wtilde S^1_1-\wtilde S^2_2\right)
\end{eqnarray}
and the total quark number is
\begin{equation}\label{2-3b}
\wtilde N= \wtilde N_1+ \wtilde N_2+ \wtilde
N_3=3\Omega+\frac{1}{2}\left(\wtilde S^1_1+\wtilde S^2_2+\wtilde
S^3_3\right).
\end{equation}
\end{subequations}
As a sub-algebra, the $su(4)$ algebra contains the $su(3)$ algebra 
which is 
generated by
\begin{equation}\label{2-4}
\wtilde S_1^2,\;\wtilde S_2^1,\;\wtilde S_2^3,\;\wtilde S_3^2,\;\wtilde
S_3^1,\; \wtilde S_1^3,\;\frac{1}{2}(\wtilde S_2^2-\wtilde
S_3^3),\;\wtilde S_1^1-\frac{1}{2} (\wtilde S_2^2+\wtilde S_3^3). 
\end{equation}
The Casimir operator ${\wtilde {\mib Q}}^2$ for the $su(3)$ algebra reads 
\begin{eqnarray}\label{2-5}
{\wtilde {\mib Q}}^2&=&
\sum_{i\neq j}{\wtilde S}_j^i{\wtilde S}_i^j
+2\left(\frac{1}{2}\left({\wtilde S}_2^2-{\wtilde S}_3^3\right)\right)^2
+\frac{2}{3}\left({\wtilde S}_1^1-\frac{1}{2}\left({\wtilde S}_2^2
+{\wtilde S}_3^3\right)\right)^2 \ .
\end{eqnarray}

The many-quark model which we investigate in this paper 
has the pairing interaction 
and particle-hole type interaction in terms of usual many-fermion system. 
The Hamiltonian is written as 
\begin{equation}\label{2-6}
{\wtilde H}_m={\wtilde H}+\chi{\wtilde {\mib Q}}^2\ , \qquad
(\chi\ : \ \hbox{\rm a\ real\ parameter})
\end{equation}
where ${\wtilde H}$ and ${\wtilde {\mib Q}}^2$ are defined by 
\begin{equation}\label{2-7}
{\wtilde H}=-G\left({\wtilde S}^1{\wtilde S}_1+{\wtilde S}^2{\wtilde S}_2
+{\wtilde S}^3{\wtilde S}_3\right) \  
\end{equation}
and Eq.(\ref{2-5}). 
We omit a kinetic term for quarks. 
Hereafter, the coupling 
constant $G$ for the pairing interaction is set to 1 without loss of 
generality. The coupling $\chi$ represents the force strength of the 
particle-hole type interaction. 
If we only have ${\wtilde H}$, namely, we take $\chi=0$, this model 
is known as the Bonn model\cite{Petry} and obeys the $su(4)$ algebra. 
Thus, we call our model (\ref{2-6}) the modified Bonn model. 
Here, since 
${\wtilde {\mib Q}}^2$ is a Casimir operator for the $su(3)$ algebra, 
then, the modified Bonn model still has a the color $su(3)$ symmetry, 
that is, 
\b\label{2-8}
[ \ {\wtilde H}_m \ , \ {\wtilde S}_j^i\ ]=0 \ , \qquad
i\ ,\ j=1,\ 2, \ 3\ .
\end{equation}

The Schwinger-type boson realization for the $su(4)$ generators 
are given as 
\begin{equation}\label{2-9}
\hat S^i=\hat a^*_i\hat b-\hat a^*\hat b_i,\quad
\hat S_i=\hat b^*\hat a_i-\hat b^*_i\hat a,\quad \hat S^j_i=(\hat
a^*_i\hat a_j-\hat b^*_j\hat b_i)+\delta_{ij}(\hat a^*\hat a-\hat
b^*\hat b),
\end{equation}
where $\hat a_i, \hat a_i^*, \hat b_i, \hat b_i^*,  \hat a,\hat
a^*,\hat b,\hat b^*\ (i=1,2,3)$ denote boson operators. 
Associated with the form (\ref{2-9}) in the boson representation, 
we define the 
$su(1,1)$ algebra: 
\begin{eqnarray}\label{2-10}
& &{\hat T}_+={\hat t}_++{\hat \tau}_+ \ , \quad
{\hat T}_-={\hat t}_-+{\hat \tau}_- \ , \qquad
{\hat T}_0={\hat t}_0+{\hat \tau}_0 \ , \nonumber\\
& &{\hat t}_+=\hat b^*\hat a^* \ , \quad
{\hat t}_-=\hat a\hat b \ , \quad
{\hat t}_0={1\over2}(\hat b^*\hat b+\hat a^*\hat a)+\frac{1}{2} \ , 
\nonumber\\
& &
{\hat \tau}_+=\sum_{i=1}^3\hat b_i^*\hat a_i^*\ , \quad
{\hat \tau}_-=\sum_{i=1}^3\hat a_i\hat b_i\ ,\quad
{\hat \tau}_0={1\over2}\sum_{i=1}^3(\hat b_i^*\hat b_i+\hat a_i^*\hat a_i)
+\frac{3}{2}\ ,
\end{eqnarray}
which satisfy 
\begin{eqnarray}
& &[\hat T_+,\hat T_-]=-2\hat T_0\ , \quad [\hat T_0,\hat T_\pm]=\hat
T_\pm \ , 
\label{2-11}\\
& &[\hat T_\mu,\hat S_i]=[\hat T_\mu,\hat S^i]=[\hat T_\mu,\hat
S_i^j]=0\ ,\quad\mu=\pm,\ 0 \ .
\label{2-12}
\end{eqnarray}
Thus, we can deal with the modified Bonn model 
in the boson space. 
The Hamiltonian (\ref{2-6}) is replaced to 
\begin{eqnarray}
& &{\hat H}_m={\hat H}+\chi{\hat {\mib Q}}^2\ , 
\label{2-13}\\
& & \ \ 
{\hat H}=-\left({\hat S}^1{\hat S}_1+{\hat S}^2{\hat S}_2
+{\hat S}^3{\hat S}_3\right) \  , \nonumber\\
& &\ \ 
{\hat {\mib Q}}^2=
\sum_{i\neq j}{\hat S}_j^i{\hat S}_i^j
+2\left(\frac{1}{2}\left({\hat S}_2^2-{\hat S}_3^3\right)\right)^2
+\frac{2}{3}\left({\hat S}_1^1-\frac{1}{2}\left({\hat S}_2^2
+{\hat S}_3^3\right)\right)^2 \ 
\label{2-14}
\end{eqnarray}
with $G=1$ for the pairing interaction. 
The relative strength of the pairing interaction to the particle-hole type 
interaction is given as $G/\chi=1/\chi$ for $G=1$.

\subsection{The exact eigenstates paying an attention to the 
pairing correlation}

The orthogonal set specified by eight quantum numbers is 
constructed in (I), namely, (I$\cdot$4$\cdot$1) as 
\begin{equation}\label{2-15}
\ket{\lambda\mu\nu\nu_0;\rho\sigma_0 TT_0}=({\hat T}_+)^{T_0-T}
{\hat Q}_+(\lambda\mu\nu\nu_0)\ket{\lambda\rho\sigma_0 T} \ . 
\end{equation}
Here, the state $\ket{\lambda\rho\sigma_0 T}$ is written as 
\beq
& &\ket{\lambda\rho\sigma_0 T}=({\hat S}^3)^{2\lambda}({\hat S}^4)^{2\rho}
\ket{\sigma_0,T} \ , \nonumber\\
& &\ \ \ket{\sigma_0,T}=\ket{m_1}=({\hat b}_1^*)^{2(\sigma_1-\sigma_0)}
({\hat b}^*)^{2\sigma_0}\ket{0} \ , 
\nonumber\\
& &\ \ \sigma_1=T-2 \  
\label{2-16}
\eeq
and the operators ${\hat S}^4$ in (\ref{2-16}) 
and ${\hat Q}_+(\lambda\mu\nu\nu_0)$ in 
(\ref{2-15}) are 
represented as 
\beq
& &
{\hat S}^4={\hat S}^1\left(
{\hat S}_1^1-\frac{1}{2}({\hat S}_2^2+{\hat S}_3^3)\right)
+({\hat S}^2{\hat S}_1^2+{\hat S}^3{\hat S}_1^3) \ , 
\label{2-17}\\
& &{\hat Q}_+(\lambda\mu\nu\nu_0)
=\sum_{\lambda_0\mu_0}\langle \lambda\lambda_0\mu\mu_0\ket{\nu\nu_0}
\sqrt{\frac{(\lambda-\lambda_0)!}{(2\lambda)!(\lambda+\lambda_0)!}}
\sqrt{\frac{(2\mu)!}{(\mu+\mu_0)!(\mu-\mu_0)!}} \nonumber\\
& &\qquad\qquad\qquad\qquad
\times({\hat S}_1^3)^{\mu+\mu_0}(-{\hat S}_1^2)^{\mu-\mu_0}
({\hat S}_2^3)^{\lambda+\lambda_0} \ .
\label{2-18}
\eeq
Further, we introduced $\Omega$-operator ${\hat \Omega}$, 
the number operators of color $i$ quarks, ${\hat N}_i$, and 
the total quark number operator ${\hat N}$ in this boson space, which have 
the eigenvalues $\Omega$, $n_i$ and $N$ acting on $\ket{m_1}$:
\begin{eqnarray}
& &\hat \Omega=n_0+\frac{1}{2}\left(\sum_{i=1}^3(\hat a^*_i
\hat a_i+\hat b^*_i\hat b_i)+\hat a^*\hat a+\hat b^*\hat
b\right)\ , 
\label{2-19}\\
& &\hat N_i=n_0+\hat a^*\hat a+(\hat b^*_i\hat b_i-\hat
a^*_i\hat a _i)+\sum_{j=1}^3\hat a^*_j\hat a_j \ ,
\label{2-20}\\
& &\hat N=3n_0+3\hat a^*\hat a+2\sum_{j=1}^3\hat a^*_j\hat
a_j+\sum_{j=1}^3\hat b^*_j\hat b_j \ , 
\label{2-21}
\end{eqnarray}
where $n_0$ implies $n_2=n_3=n_0$ in the state $\ket{m_1}$. 
We have
\beq\label{2-22}
& &n_0=\Omega-\sigma_1 \ , \qquad n_1=\Omega+\sigma_1-2\sigma_0  \ , 
\nonumber\\
& &{\rm i.e.,}\ \ 
n_1-n_0=2(\sigma_1-\sigma_0) \ .
\eeq

Hereafter, we take the particle picture developed in (I). 
The results with respect to the hole picture 
are the similar to those of the particle picture. 
The eigenvalue equation for ${\hat H}_m$ and the eigenvalue are 
obtained as
\beq
& &{\hat H}_m\ket{\lambda\mu\nu\nu_0;\rho\sigma_0 TT_0}
=E_{\sigma_1\sigma_0\rho\lambda}^{(m)}
\ket{\lambda\mu\nu\nu_0;\rho\sigma_0 TT_0}\ , 
\nonumber\\
& &E_{\sigma_1\sigma_0\rho\lambda}^{(m)}
=E_{\sigma_1\sigma_0\rho\lambda}+\chi
F_{\sigma_1\sigma_0\rho\lambda}^{(p)} \ , 
\label{2-23}\\
& &\ \ 
E_{\sigma_1\sigma_0\rho\lambda}
=
-\left(2\lambda(2\sigma_0+1-2\rho-2\lambda)
+2\rho(2\sigma_1+3-2\rho)\right)\nonumber\\
& &\qquad\qquad\ 
=
-\left(\frac{1}{2}N-\frac{1}{2}(2n_0+n_1)-2\rho\right)
\left(2\Omega+1-\frac{1}{2}n_1-\frac{1}{2}N\right)\nonumber\\
& &\qquad\qquad\quad\ 
-2\rho(2\Omega+3-2n_0-2\rho) \nonumber\\
& &\qquad\qquad\ 
=E_{Nn_0n_1\rho} \ , 
\nonumber\\
& &\ \ 
F_{\sigma_1\sigma_0\rho\lambda}^{(p)}=2\lambda(\lambda+1)
+\frac{2}{3}\Bigl(2(\sigma_1-\sigma_0)-2\rho+\lambda\Bigl)\Bigl(
2(\sigma_1-\sigma_0)-2\rho+\lambda+3\Bigl)\nonumber\\
& &\qquad\qquad\ 
=G_{Nn_0n_1}+2E_{\sigma_1\sigma_0\rho\lambda} \ , \nonumber\\
& &\ \ \ \ 
G_{Nn_0n_1}=2(\Omega-n_0)(\Omega-n_0+3)+(\Omega-n_1)^2
-\frac{1}{3}(3\Omega-N)(3\Omega-N+6) \ . \nonumber\\
& &\label{2-24}
\eeq
Here, we used the following relations for $\Omega$ and $N$ obtained 
when the operators ${\hat \Omega}$ and ${\hat N}$ in Eqs.(\ref{2-19}) and 
(\ref{2-21}) act on the state $\ket{\lambda\rho\sigma_0\sigma_1}$ 
in Eq.(\ref{2-16}): 
\beq\label{2-25}
\Omega&=&n_0-2+T=n_0+\sigma_1 \ , 
\nonumber\\
N&=&3n_0+4(\lambda+\rho)+2\sigma_1-2\sigma_0 \nonumber\\
&=&2n_0+n_1+4(\lambda+\rho) \ . 
\eeq
Here, we further used the relation (\ref{2-16}), that is, 
$T=\sigma_1+2$.

\subsection{Triplet formation}

Next, we introduce another form for the orthogonal set. 
This state corresponds to the triplet formation, in which 
the fermion number changes in unit 3 as was mentioned in 
Eq.(I$\cdot$3$\cdot$38).

The energy eigenstates are constructed by 
\beq\label{2-26}
\kkett{\lambda\mu\nu\nu_0;t\tau T T_0}
&=&({\hat T}_+)^{T_0-T}
\kkett{\lambda\mu\nu\nu_0;t\tau T}=({\hat T}_+)^{T_0-T}
{\hat Q}_+(\lambda\mu\nu\nu_0)\kkett{\lambda\tau tT} \ , \nonumber\\
\kkett{\lambda\mu\nu\nu_0;t\tau T}
&=&({\hat O}_+(t\tau))^{T-(t+\tau)}
{\hat Q}_+(\lambda\mu\nu\nu_0)\kket{\lambda\tau}\otimes\kett{t}
\nonumber\\
&=&\sum_{t_0\tau_0; t_0+\tau_0=T_0}\!\!\!\!\!\!{}'\ \ 
\frac{(-1)^{\tau_0-\tau}\Gamma(T-(t+\tau)+1)\Gamma(2t)\Gamma(2\tau)}
{\Gamma(t_0-t+1)\Gamma(\tau_0-\tau+1)\Gamma(t_0+t)\Gamma(\tau_0+\tau)}
\nonumber\\
& &\qquad\qquad\qquad
\times
\kket{\lambda\mu\nu\nu_0;\tau\tau_0}\otimes\kett{tt_0} ,\nonumber\\
\kkett{\lambda\tau tT}&=&({\hat O}_+(t\tau))^{T-(t+\tau)}
\kket{\lambda\tau}\otimes\kett{t} \ , 
\eeq
where the operator ${\hat Q}_+(\lambda\mu\nu\nu_0)$ is defined in 
Eq.(\ref{2-18}) and ${\hat O}_+$ is introduced as 
\beq\label{2-27}
& &
{\hat O}_+(t\tau)={\hat t}_+({\hat t_0}+t+\epsilon)^{-1}
-{\hat \tau}_+({\hat \tau}_0+\tau+\epsilon)^{-1} \ .
\eeq
The states appearing in (\ref{2-26}) are defined as 
\beq\label{2-28}
& &\kket{\lambda\tau}=({\hat a}_3^*)^{2\lambda}
({\hat b}_1^*)^{2\tau-3-2\lambda}\ket{0}\ , \nonumber\\
& &\kett{t}=({\hat b}^*)^{2t-1}\ket{0} \ , \nonumber\\
& &\kket{\lambda\mu\nu\nu_0;\tau\tau_0}
={\hat Q}_+(\lambda\mu\nu\nu_0)({\hat \tau}_+)^{\tau_0-\tau}
\kket{\lambda\tau} \ , \nonumber\\
& &\kett{tt_0}=({\hat t}_+)^{t_0-t}\kett{t} \ . 
\eeq
Then, the eigenvalue equation and energy eigenvalue are derived as 
\begin{eqnarray}
{\hat H}_m\kkett{\lambda\mu\nu\nu_0;t\tau T T_0}
&=&E_{Tt\tau\lambda}^{(m)}\kkett{\lambda\mu\nu\nu_0;t\tau T T_0} \ , 
\label{2-29}\\
E_{Tt\tau\lambda}^{(m)}&=&
E_{Tt\tau\lambda}+\chi F_{\tau\lambda}^{(t)} \ , \nonumber\\
E_{Tt\tau\lambda}&=&
-(T-t-\tau)(T+t+\tau-1)-4t\lambda \ , \nonumber\\
F_{\tau\lambda}^{(t)}&=&
2\lambda(\lambda+1)+\frac{2}{3}((2\tau-3)-\lambda)
((2\tau-3)-\lambda+3)\ .
\label{2-30}
\end{eqnarray}
Operation of ${\hat \Omega}$ and ${\hat N}$ on $\kkett{\lambda\tau tT}$ 
leads to 
\begin{eqnarray}\label{2-31}
& &{T}={\Omega}+2-n_0 \ , 
\nonumber\\
& &{t}={\Omega}+1-\frac{1}{3}({\tau}-2\lambda)-\frac{1}{3}{N}
\ . 
\end{eqnarray}
Thus, $E_{Tt\tau\lambda}$ is rewritten as 
\begin{eqnarray}\label{2-32}
E_{Tt\tau\lambda}&=&
-\left(\frac{1}{3}N-n_0-\frac{1}{3}(2\tau-3)-\frac{2}{3}\lambda\right) 
\nonumber\\
& &\times \left(2\Omega+3-\frac{1}{3}N-n_0+\frac{1}{3}(2\tau-3)+
\frac{2}{3}\lambda\right)
\nonumber\\
& &-4\lambda\left(\Omega+\frac{1}{2}-\frac{1}{3}N-\frac{1}{6}
(2\tau-3)+\frac{2}{3}\lambda\right) \ .
\end{eqnarray}

\section{Physical interpretation of the eigenstates}

In this section, the physical interpretation of the various states 
is investigated for the Hamiltonian ${\hat H}$, namely, we 
put $\chi=0$ in Eq.(\ref{2-13}). 

\subsection{Pairing correlation}

\subsubsection{Correspondence to the original fermion states}

From (\ref{2-25}) for the quark number in the pairing state (\ref{2-16}), 
we can obtain the following relation: 
\b\label{3-1}
2\lambda=(N-(2n_0+n_1))/2-2\rho \ . 
\end{equation}
From the construction of the state, $2\lambda$ and $2\rho$ 
must be positive integers. 
Therefore, $N-(2n_0+n_1)$ have to 
be a positive even integer. 
This means that the change in the fermion number relatively to that 
of the minimum weight state is restricted to even number, i.e., 
it is of the pairing-type. 
From (\ref{2-22}), the number of quarks in the state $\ket{m_1}$ is given as 
\b\label{3-2}
N=n_1+n_2+n_3=n_1+2n_0 \ , \quad (n_2=n_3=n_0) \ 
\end{equation}
and each quark number with color $i$ is also written as 
\begin{eqnarray}\label{3-3}
\ket{m_1}\ : \  \Biggl\{
\begin{array}{@{\,}c@{\,}}
{\rm numbers\ of\ color}\ 1 = n_1\\
{\rm numbers\ of\ color}\ 2 = n_0\\
{\rm numbers\ of\ color}\ 3 = n_0
\end{array}
\end{eqnarray}

In order to investigate the physical interpretation of the exact eigenstates 
in the boson representation, we return to the original fermion 
states which correspond to the boson states in the 
boson representation. 
In the original fermion space, the operator ${\hat S}^4$ 
has the following correspondence: 
\begin{eqnarray}\label{3-4}
{\hat S}^4 \longleftrightarrow 
{\tilde S}^4&=&
\frac{1}{2}\sum_{m}c_{2m}^*c_{3{\tilde m}}^*
\sum_{m}(c_{2m}^*c_{2m}+c_{3m}^*c_{3m}-2c_{1m}^*c_{1m})\nonumber
\\
& &+\sum_{mm'}(c_{3m}^*c_{1{\tilde m}}^*c_{2m'}^*c_{1m'}
+c_{1m}^*c_{2{\tilde m}}^*c_{3m}^*c_{1m}) \ .
\end{eqnarray}
Namely, the operator ${\hat S}^4$ creates two quarks whose colors 
are 2 and 3. 
Further, the operators ${\hat S}^3$, ${\hat S}_2^3$, ${\hat S}_1^2$ 
and ${\hat S}_1^3$ have the correspondences in the following manner: 
\begin{eqnarray}\label{3-5}
{\hat S}^3 \longleftrightarrow 
{\tilde S}^3 &=&
\sum_{m}c_{1m}^*c_{2{\tilde m}}^* \ , \nonumber\\
{\hat S}_2^3 \longleftrightarrow 
{\tilde S}_2^3&=&\sum_m c_{3m}^*c_{2m} \ , \nonumber\\
{\hat S}_1^2 \longleftrightarrow 
{\tilde S}_1^2&=&\sum_{m}c_{2m}^*c_{1m} \ , \nonumber\\
{\hat S}_1^3 \longleftrightarrow 
{\tilde S}_1^3&=&\sum_{m}c_{3m}^*c_{1m} \ .
\end{eqnarray}
Thus, ${\hat S}^3$ creates two quarks with color 1 and 2, 
${\hat S}_2^3$, ${\hat S}_1^2$ 
and ${\hat S}_1^3$ create one quark with color 3, 2 and 3 
and annihilate one quark with color 2, 1 and 1, respectively, 
in terms of the original fermion space. 
Thus, the state (\ref{2-15}) under $T=T_0$ includes the following 
quarks: 
\begin{eqnarray}\label{3-6}
\begin{array}{@{\,}ll@{\,}}
{\rm numbers\ of\ color}\ 1 & = n_1+2\lambda-2\mu \ , \\
{\rm numbers\ of\ color}\ 2 & =n_0+2\rho+\lambda+\lambda_0+\mu-\mu_0 \ ,  \\
{\rm numbers\ of\ color}\ 3 & = n_0+2\rho+\lambda+\lambda_0+\mu+\mu_0 \ .
\end{array}
\end{eqnarray}

From (\ref{2-8}), the Hamiltonian commutes with ${\hat S}_1^2$, 
${\hat S}_1^3$ and ${\hat S}_2^3$. Thus, if $T=T_0$, 
the energy eigenvalue is determined by the state 
$\ket{\lambda\rho\sigma_0 T}$ in Eq.(\ref{2-16}). 
Hereafter, we use the notation $E_{Nn_0n_1\rho}$ given in Eq.(\ref{2-24}).

\subsubsection{Color superconducting state where colors 2 and 3 are coupled}

For the state $\ket{\lambda\rho\sigma_0 T}$ in Eq.(\ref{2-16}), 
let us adopt the following parameterization: 
\begin{eqnarray}\label{3-8}
& &n_0=0\ , \qquad n_1=2\Omega \, \qquad 2\rho=q' \nonumber\\
& &\mu=\mu_0=0\ , \qquad \lambda=\lambda_0=0 \ . 
\end{eqnarray}
Then, as is seen from Eq.(\ref{3-6}), the state (\ref{2-16}) 
under the above conditions includes 
the quarks as 
\begin{eqnarray}\label{3-9}
\begin{array}{@{\,}l@{\,}}
{\rm numbers\ of\ color}\ 1 = n_1\ (=2\Omega) \ , \\
{\rm numbers\ of\ color}\ 2 = 2\rho\ (=q') \ , \\
{\rm numbers\ of\ color}\ 3 = 2\rho\ (=q') \ . 
\end{array}
\end{eqnarray}
Here, the total quark number $N$ is written as 
\begin{equation}\label{3-10}
N=n_1+2\rho+2\rho =2\Omega+2q'\ .
\end{equation}
Thus, in the case $\chi=0$, 
the energy eigenvalue (\ref{2-24}) with $\chi=0$ is recast into 
\begin{eqnarray}\label{3-11}
E_{N\ n_0=0\ n_1=2\Omega\ \rho=q'/2}&=&
-2\rho(2\Omega +3-2\rho)\nonumber\\
&=&-q'(2\Omega+3-q') \ .
\end{eqnarray}

In the original fermion system, the following state 
can be constructed:
\begin{eqnarray}\label{3-12}
\ket{\Psi(q',\Omega)}=({\tilde S}^1)^{q'}\cdot
\prod_{m=1}^\Omega c_{1m}^*c_{1{\tilde m}}^*\ket{0} \ .
\end{eqnarray}
This state is nothing but the color superconducting state 
in which the quarks with color 2 and 3 
are coupled. 
The number of quarks with each color in this state is as follows: 
\begin{eqnarray}\label{3-13}
& &
\begin{array}{@{\,}l@{\,}}
{\rm numbers\ of\ color}\ 1 = 2\Omega \ , \\
{\rm numbers\ of\ color}\ 2 = q' \ , \\
{\rm numbers\ of\ color}\ 3 = q' \ 
\end{array}
\end{eqnarray}
and the total quark number $N$ is obtained as 
\begin{eqnarray}\label{3-14}
& &N=2\Omega+2q' \ .
\end{eqnarray}
For this state, the energy eigenvalue is derived easily in the fermion space 
and the result 
is as follows: 
\begin{eqnarray}\label{3-15}
E_N=-q'(2\Omega+3-q') \ .
\end{eqnarray}
Thus, we conclude that the state described 
in the boson space, Eq.(\ref{2-16}), under the conditions (\ref{3-8}) 
corresponds to the color superconducting state in the original 
fermion space, in which the quarks with color 2 and 3 are coupled.

\subsection{triplet formation}

As was discussed in (I), the operators ${\hat b}^*$, ${\hat a}^*$, 
$\{{\hat t}_\mu\}$ and $\{{\hat \tau}_{\mu}\}$ commute with the $su(3)$ 
generators. Namely, they are invariant under the group $SU(3)$. 
Further, from Eq.(I$\cdot$5$\cdot$17), 
we derived the following relation in the state (\ref{2-26}) with 
(\ref{2-28}): 
\b\label{3-7}
\lambda=\mu=\nu=\nu_0=0 \quad {\rm for} \quad \tau=\frac{3}{2} \ . 
\end{equation}
Thus, it is seen that under $T=T_0$, the state (\ref{2-26}) with $\tau=3/2$ 
is color neutral. 
Of course, if $\tau > 3/2$, the state is colored one. 

As is similar to the case of the pairing correlation, 
the energy eigenvalue in the case of triplet formation 
is determined by the state $\kkett{\lambda\tau tT_0}$ 
when $T=T_0$. 
Hereafter, we restricted ourselves to the case $T=T_0$.

\subsection{Relation of the states with 
pairing correlation to those with triplet formation}

In our previous paper (II), 
the state $\kket{lsrw}$ is introduced and is defined as 
\b\label{3-16}
\kket{lsrw}=({\hat S}^3)^{2l}({\hat q}^1)^{2s}({\hat B}^*)^{2r}
({\hat b}^*)^{2w}\ket{0} \ , 
\end{equation}
where the newly introduced operators ${\hat q}^i$ and ${\hat B}^*$ 
are defined as 
\beq\label{3-17}
& &{\hat q}^i={\hat b}_i^*{\hat b}-{\hat a}^*{\hat a}_i \ , \qquad
{\hat B}^*=\sum_i{\hat S}^i{\hat q}^i \ . 
\eeq
Then, the state with pairing correlation, 
$\ket{\lambda\rho\sigma_0T}$ in Eq.(\ref{2-16}), 
can be recast into 
\beq\label{3-18}
\ket{\lambda\rho\sigma_0\sigma_1}
&=&\kket{\lambda,\sigma_1-\sigma_0-\rho,\rho,\sigma_1} \ 
\eeq
with $\sigma_1=T-2$. (See, (II$\cdot$5$\cdot$11).)

In the state (\ref{2-15}) with $T=T_0$  and 
\begin{eqnarray}\label{3-19}
& &\mu=\mu_0=0 \ , \qquad \lambda=\lambda_0=0 \ , \qquad n_0=0 \ , \qquad 
n_1=2\rho \ , \nonumber\\
& &\qquad\qquad\qquad\qquad\qquad\qquad\qquad\qquad
\qquad (\sigma_1-\sigma_0=\frac{1}{2}(n_1-n_0)) \ , 
\end{eqnarray}
the numbers of quarks with color $i$ are written from (\ref{3-6}) as 
\begin{eqnarray}\label{3-20}
& &
\begin{array}{@{\,}l@{\,}}
{\rm numbers\ of\ color}\ 1 = n_1 \ , \\
{\rm numbers\ of\ color}\ 2 = 2\rho=n_1 \ , \\
{\rm numbers\ of\ color}\ 3 = 2\rho=n_1 \ , 
\end{array}\\
& &{\rm i.e.,}\qquad 2\rho=n_1=\frac{N}{3} \ . \nonumber
\end{eqnarray}
Therefore, this state is color neutral. 
Here, the state (\ref{3-18}) is written as 
\b\label{3-21}
\kket{00N/6\ T-2} \ . 
\end{equation}
For this state, the energy eigenvalue (\ref{2-23}) with $\chi=0$ 
is obtained as 
\begin{eqnarray}\label{3-22}
E_{N\ \rho=N/6\ n_0=0\ n_1=N/3}&=&
-\frac{N}{3}\left(2\Omega+3-\frac{N}{3}\right) \ . 
\end{eqnarray}
It will be shown in the following that 
this energy eigenvalue is identical with the one derived by the 
state which forms the color neutral 
quark triplet, namely $\kkett{\lambda\tau tT}$ 
in Eq.(\ref{2-26}). 
From (\ref{2-32}) with $n_0=0$ and $\tau=3/2$, 
which means $\lambda=0$ from Eq.(\ref{3-7}), the energy eigenvalue 
in the case of the triplet formation is written as 
\begin{eqnarray}\label{3-23}
E_{T=\Omega+2\ t=\Omega+1/2+N/3\ \tau=3/2\ \lambda=0}&=&
-\frac{N}{3}\left(2\Omega+3-\frac{N}{3}\right) \ .
\end{eqnarray}
Of course, $\tau=3/2$ reveals that this state is the color neutral state. 
As was shown in (II), the state with triplet formation 
$\kkett{\lambda\tau tT}$ in Eq.(\ref{2-26}) is nothing but the 
state $\kket{lsrw}$ in Eq.(\ref{3-16}) with the following relations:
\b\label{3-24}
l=\lambda\ , \quad
s=\frac{1}{2}(2\tau-3-2\lambda) \ , \quad
r=\frac{1}{2}(T-t-\tau) \ , \quad
w=T-2\ , 
\end{equation}
which was given in (II$\cdot$5$\cdot$25). 
Thus, the above state with 
$n_0=0$ and $\tau=3/2$ has the following parameters as 
\b\label{3-25}
l=0\ , \quad
s=0 \ , \quad
r=\frac{1}{2}\left(T-t-\frac{3}{2}\right)=\frac{N}{6} \ , \quad
w=T-2\ , 
\end{equation}
where we used the relation in Eq.(\ref{2-31}). 
Thus, this state is written as 
\b\label{3-26}
\kket{00N/6\ T-2} \ . 
\end{equation}
It should be noted that this triplet state is identical with the state 
with pairing correlation in Eq.(\ref{3-21}).

\section{Ground state for $\chi \neq 0$}

In this section, 
the main aim of this paper is treated, namely, 
we investigate the ground state in our 
modified Bonn model whose Hamiltonian is given in Eq.(\ref{2-13}). 
In the previous paper (I), we investigated the energy minimum state 
and its energy eigenvalue in the pairing correlation and triplet formation, 
respectively.  

In the case of triplet formation, the energy of 
the colored state with $\tau\neq 3/2$ increases together with the 
particle-hole type 
coupling strength $\chi$ increasing. 
Thus, if the coupling strength $\chi$ is greater than a certain 
value, which was given in Eq.(I$\cdot$5$\cdot$25) or (I$\cdot$5$\cdot$26) 
for the particle or hole picture, respectively, 
the energy of the colored state 
is larger than that of the color neutral triplet state 
with $\tau=3/2$ because the energy of the color neutral triplet 
has no change with respect to $\chi$. 
Thus, the color neutrality is retained in our model. 

As is similar to the case of the triplet formation, in the case 
of the pairing correlation in which the eigenstate is 
given in Eq.(\ref{2-15}) or (\ref{2-16}), 
the energy of the colored states 
is pushed up when the coupling strength $\chi$ increases.

Here, let us search the ground state in the various coupling 
strength $\chi$ numerically. 
Hereafter, we set $n_0=0$ whose parameter was used in (I). 
In (I), the energy minimum states were obtained in both the case 
of pairing correlation and triplet formation, respectively. 
As for the triplet formation, the energy is given in (\ref{3-23}) 
for $\chi=0$. 
As for the pairing correlation, the minimum energy states and 
minimum energy ${\cal E}$ are determined analytically for various $\chi$. 
We summarize them: 
\beq\label{4-1}
& &{\rm (1)}\ -\frac{1}{6}\cdot\frac{\Omega+6}{\Omega+2}
\leq \chi \leq -\frac{1}{6}\nonumber\\
& &\qquad\qquad
 {\rm for}\ \ 0 \leq N \leq \frac{6(2\Omega-3-6\chi)}{5+6\chi}
\nonumber\\
& &\qquad\qquad\qquad\quad
{\cal E}=E(N)+F(N)\ , \nonumber\\
& &\qquad\qquad\qquad\qquad\quad
E(N)=-\frac{1}{4}N(4\Omega+2-N) \ , \quad
F(N)=\frac{\chi}{6}N(N+6) \ ,
\nonumber\\
& &\qquad\qquad\qquad\quad
n_1=0 \ , \nonumber\\
& &\qquad\qquad 
{\rm for}\ \frac{6(2\Omega-3-6\chi)}{5+6\chi} < N \leq 3\Omega
\nonumber\\
& &\qquad\qquad\qquad\quad
{\cal E}=E(N)+F(N)
-\frac{N}{36}[(5N-12\Omega+18)+6\chi(N+6)] \ , \nonumber\\
& &\qquad\qquad\qquad\quad
n_1=\frac{N}{3} \ , \nonumber\\
& &{\rm (2)}\ -\frac{1}{6} < \chi \leq \frac{1}{6}(2\Omega-3)\nonumber\\
& &\qquad\qquad 
{\rm for}\ 0 \leq N \leq \frac{2\Omega}{1+2\chi}-3
\nonumber\\
& &\qquad\qquad\qquad\quad
{\cal E}=E(N)+F(N) \ , \nonumber\\
& &\qquad\qquad\qquad\quad 
n_1=0 \ , \nonumber\\
& &\qquad\qquad 
{\rm for}\ \frac{2\Omega}{1+2\chi}-3 < N < 3\Omega-\frac{9}{2}-9\chi
\nonumber\\
& &\qquad\quad\qquad\qquad
{\cal E}=E(N)+F(N)
-\frac{1}{4(1+6\chi)}\left[
(N-2\Omega+3)+2\chi(N+3)\right]^2 \ , \nonumber\\
& &\qquad\qquad\qquad\quad 
n_1=\frac{(N-2\Omega+3)+2\chi(N+3)}{1+6\chi} \ , \nonumber\\
& &\qquad\qquad 
{\rm for}\ 3\Omega-\frac{9}{2}-9\chi \leq N \leq 3\Omega
\nonumber\\
& &\qquad\qquad\qquad\quad
{\cal E}=E(N)+F(N)
-\frac{N}{36}[(5N-12\Omega+18)+6\chi(N+6)] \ , \nonumber\\
& &\qquad\qquad\qquad\quad 
n_1=\frac{N}{3} \ , \nonumber\\
& &{\rm (3)}\ \chi \geq \frac{1}{6}(2\Omega-3)\nonumber\\
& &\qquad\qquad 
{\rm for}\ 0 \leq N \leq 3\Omega
\nonumber\\
& &\qquad\qquad\qquad\quad
{\cal E}=E(N)+F(N)
-\frac{N}{36}[(5N-12\Omega+18)+6\chi(N+6)] \ , \nonumber\\
& &\qquad\qquad\qquad\quad 
n_1=\frac{N}{3} \ .
\eeq
\begin{figure}[b]
\begin{center}
\includegraphics[height=5.0cm]{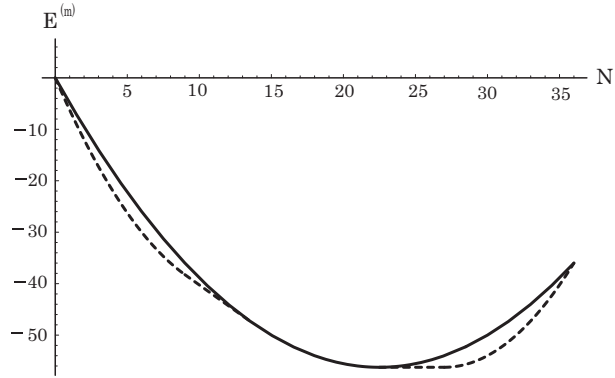}
\caption{The total energies are shown as a function of the particle 
number $N$ with $\Omega=6$ in the case of $\chi=0$. 
The solid curve represents the minimum 
energy 
calculated in the case of the triplet formation. 
The dashed curve represents the minimum energy calculated in the case of 
the pairing correlation. 
}
\label{fig:4-1}
\end{center}
\end{figure}
In the case of the range $3\Omega \leq N \leq 6\Omega$, 
$N$ appearing in all the terms except $E(N)$ should be replaced by 
$6\Omega-N$ and as for $E(N)$, the following should be used: 
\beq\label{4-2}
E(N)=-\frac{1}{4}(N-2\Omega)(6\Omega+6-N) \ .
\eeq

Here, we omit the case $\chi<-(\Omega+6)/(6(\Omega+2))$ because 
we treat the model based on the original Bonn model which 
has the attractive pairing type interaction with $\chi=0$. 
If $\chi$ is negative with large absolute value, 
the model mainly reveals the nature with 
the attractive particle-hole type interaction. 
Namely, this model gives a system with 
the attractive particle-hole type interaction plus relatively small attractive 
pairing interaction. 
Our purpose in this paper is to investigate the Bonn model with 
a modification. 
Therefore, the above-mentioned situation with large negative $\chi$ should be 
discussed in another context.

Our next task is to compare the energy of the state with 
pairing correlation in Eq.(\ref{2-16}) with that with triplet 
formation in Eq.(\ref{2-26}). 
In Fig.\ref{fig:4-1}, 
the total energies are shown as a function of the particle 
number $N$ in the case $\chi=0$ with $\Omega=6$. 
The solid curve represents the minimum 
energy obtained in the case of the color neutral triplet formation. 
The dashed curve represents the minimum energy obtained in the case of 
the pairing correlation. 
It is shown that, partially, 
the energy of the colored pairing state is lower than that of the color 
neutral triplet state. 
\begin{figure}[t]
\begin{center}
\includegraphics[height=5.0cm]{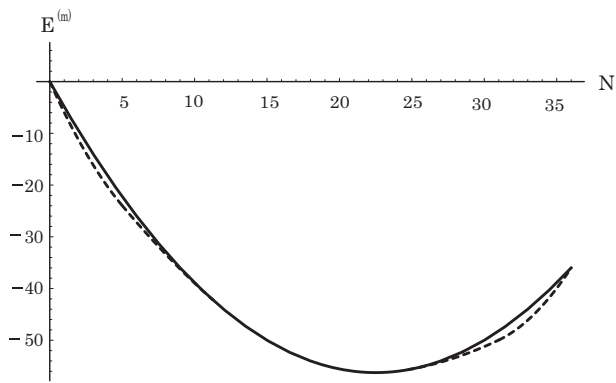}
\caption{The total energies are shown as a function of the particle 
number $N$ with $\Omega=6$ in the case of $\chi=1/4$. 
The solid curve represents the minimum 
energy 
calculated in the case of the triplet formation. 
The dashed curve represents the minimum energy calculated in the case of 
the pairing correlation. 
}
\label{fig:4-2}
\end{center}
\end{figure}
~
\begin{figure}[t]
\begin{center}
\includegraphics[height=5.0cm]{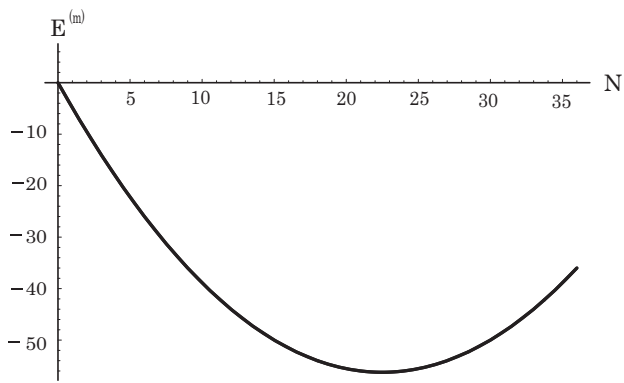}
\caption{The total energies are shown as a function of the particle 
number $N$ with $\Omega=6$ in the case of $\chi=3/2$. 
The solid curve represents the minimum 
energy calculated in the case of the triplet formation. 
This curve is identical with the one which represents 
the minimum energy calculated in the case of 
the pairing correlation. 
}
\label{fig:4-3}
\end{center}
\end{figure}
If we introduce the $su(4)$ symmetry broken but color $su(3)$ symmetric 
term $\chi{\hat {\mib Q}}^2$ in the original Bonn model, the situation is 
modified. In Fig.\ref{fig:4-2}, the same results are depicted 
except for the coupling strength of the particle-hole type $\chi$ 
which we take $\chi=1/4$. 
The energy of the colored state with pairing correlation is pushed up 
and this energy is close to that of the color neutral triplet formation. 
Finally, when the coupling strength is equal to $\chi=3/2$ or 
greater than 3/2 in the case $\Omega=6$, 
the energy with pairing correlation is equivalent to that of 
the color neutral triplet, which is described in \S 3.3. 
In this region of the coupling strength, namely, $\chi\geq 3/2$ with 
$\Omega=6$, the variable $n_1$ is equal to $N/3$ in the state with pairing 
correlation. 
Thus, the state with pairing correlation is reduced to the state 
with color neutral triplet as is mentioned in Eqs. (\ref{3-21}) and 
(\ref{3-26}). The calculated energies are shown in Fig.\ref{fig:4-3}.

\begin{figure}[t]
\begin{center}
\includegraphics[height=5.5cm]{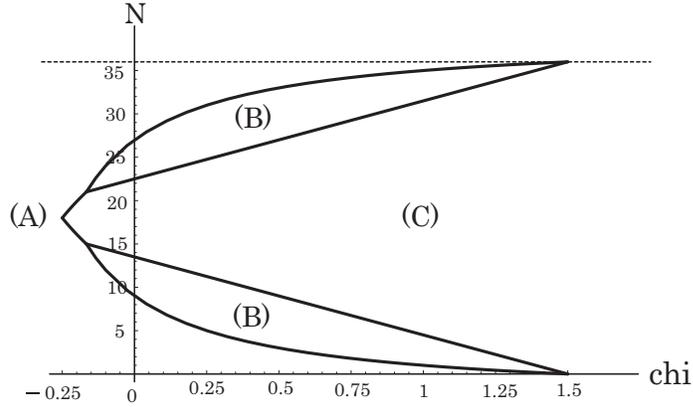}
\caption{The phase diagram is shown with respect to the particle number $N$ 
(vertical axis) and the coupling constant $\chi$ (horizontal axis). 
The parameter $\Omega$ is taken as 6, so the physical 
region is determined by the relation $0 \leq N \leq 6\Omega=36$. 
The region (A) represents the one with $n_1=0$ and the states in this 
area are pairing states in general. 
The region (C) represents the one with $n_1=N/3$. Thus, the states in 
this area are color neutral with triplet formation. 
The region (B) represents the transition region in which $0<n_1<N/3$. 
}
\label{fig:4-4}
\end{center}
\end{figure}

As was investigated in detail in (I), 
in the case of the state with pairing correlation in Eq.(\ref{2-16}), 
the variable $n_1$ plays a role of the order parameter for the 
phase transition from pairing state to color neutral triplet state. 
If $n_1=N/3$, the state is equivalent to the state with triplet formation. 
Thus, according to the value of $n_1$, 
we can give the phase diagram 
based on Eqs.(\ref{4-1}) and (\ref{4-2}) 
as is seen in Fig.\ref{fig:4-4}, in which 
the phase diagram is shown with respect to the particle number $N$ 
(vertical axis) and the coupling strength $\chi$ (horizontal axis). 
The variables are fixed as $\Omega=6$ and $n_0=0$, so the physical 
region is determined by the relation $0 \leq N \leq 6\Omega=36$. 
The region (A) represents the one with $n_1=0$ and the states in this 
area are color-pairing states in general. 
The region (C) represents the one with $n_1=N/3$. Thus, the states in 
this area are color neutral with triplet formation. 
The region (B) represents the transition region in which $0<n_1<N/3$. 
In the region (B), the order parameter $n_1$ is changed continuously 
with respect to the particle number $N$ under fixed $\chi$. 
Thus, the structure of the ground state is changed and 
it may be allowed to say that the phase change occurs. 

\section{Discussions and concluding remarks}

In the original Bonn model, that is, $\chi=0$ in Eq.(\ref{2-13}), 
the only pairing interaction between quarks with different colors 
is included, which may be regarded as an effective model of QCD. 
However, it is known that, in this model, the colored state such as the 
color superconducting state is energetically favorable. 
This fact leads to the pairing instability any time and to breaking 
of the color neutrality in the ground state, while this model reveals 
a nice feature that this model leads to the quark-triplet formation as a 
nucleon without the three-body correlation. 
As was pointed out in (I), the original Bonn model has the 
$su(4)$ symmetry. 
Thus, the room in which the $su(4)$ symmetry is broken but the color 
$su(3)$ symmetry is still retained is rest. 
Therefore, we introduced a particle-hole type interaction, 
$\chi{\hat {\mib Q}}^2$, in the previous paper (I), where 
${\hat {\mib Q}}^2$ is the Casimir operator of the $su(3)$ sub-algebra.

In this effective model of QCD, we can control the strength of pairing 
interaction and the strength of particle-hole type interaction 
which are represented by $G$ and $\chi$, respectively. 
Here, we take the ratio $R=G/\chi$, in which we fix $G=1$ in this paper. 
If the ratio $R$ is small, for example, $R\leq 2/3$ under 
the restriction $\Omega=6$, the ground state is the color neutral 
triplet state in general. 
However, when $R$ increases from $2/3$, the ground state is 
no longer the color neutral quark triplet state under a specific 
value of the particle number $N$. 
Namely, the color pairing state becomes energetically stable. 
This fact indicates that the color pairing instability is realized 
in a certain region of $N$ when the strength of the pairing 
interaction, $G$, is relatively large for the strength of the particle-hole 
type interaction, $\chi$, that is, $R$ goes beyond a certain critical value. 
If it is allowed to use the terminology of the infinite nuclear 
and/or quark matter, this situation developed in our papers is resemble to 
the phase transition between the color superconducting and the normal 
nuclear phases. 
Of course, since we cannot deduce the interaction terms of the pairing and 
the particle-hole types from QCD directly, 
the parameter $R$ is unknown in this stage. 
However, it may be a possible scenario that $R$ depends on 
the baryon density in which $R$ may increases as the baryon density 
increases. As a result, the pairing instability may occur at high 
baryon density.

\section*{Acknowledgement} 
One of the authors (Y.T.) 
is partially supported by the Grants-in-Aid of the Scientific Research 
No. 18540278 from the Ministry of Education, Culture, Sports, Science and 
Technology in Japan.


\end{document}